\documentclass[aps, prb, 12pt, reprint]{revtex4-1}

\usepackage{graphicx}

\begin{document}

\begin{abstract}
In its most general form, a `secret objective' is any inconsistency between the experimental reality and the information provided to students prior to starting work on an experiment. Students are challenged to identify the secret objectives and then given freedom to explore and understand the experiment, thus encouraging and facilitating genuine inquiry elements in introductory laboratory courses. Damping of a simple pendulum is used as a concrete example to demonstrate how secret objectives can be included. We also discuss the implications of the secret objectives method and how this can provide a link between the concepts of problem based learning and inquiry style labs.
\end{abstract}

\title{Secret objectives: promoting inquiry and tackling preconceptions in teaching laboratories}

\author{Paul A. Bartlett}
\email{paul.bartlett@ucl.ac.uk} %
\affiliation{Department of Physics and Astronomy, University College London, WC1E 6BT, UK}
\author{K. Dunnett}
\email{kirsty.dunnett@su.se}
\affiliation{Nordita, KTH Royal Institute of Technology and Stockholm University, Roslagstullsbacken 23, SE-106 91 Stockholm, Sweden}
\date{\today}

\maketitle

\section{Introduction}	

An underlying assumption of much practical work in physics courses is that it provides students with opportunities to develop practical skills (e.g.: equipment use) or reinforces knowledge by allowing students to interact directly with at least one physical realisation of the theoretical concept(s) \cite{AJP.40.584_Reichert, RevEdRes.52.201, SciEd.88.28}. In the former case, the introduction unfamiliar equipment often leads to very precise scripts that students follow slavishly to minimise problems they encounter, while the latter type of laboratories may provide less procedural details, but usually lead to previously known results \cite{AJP.40.584_Reichert}. Laboratory work with the express purpose of reinforcing lecture content is usually very precisely scripted, with very little scope for students to deviate from the instructions, and the aim is usually to obtain a known result \cite{JColSciTeach.38.52}. These `cookbook' style laboratory tasks have long been criticised for the lack of critical and independent thought that students need in order to complete them \cite{SciEd.88.28, ThePhysTeach.53.6.349}. More recently, the effect of different styles of laboratory work on the development of expert-like beliefs has been studied in some detail, with the conclusion that laboratory courses that allow for some element of independent inquiry promote the development of expert-like beliefs \cite{PRPER.12.020132}. On the other hand, very precisely scripted laboratory work, where the students are merely to demonstrate a particular phenomenon met in lectures not only have `no added value' \cite{PRPER.13.010129}, but may actually lead to less-expert beliefs about the nature of experimentation in physics and science in general \cite{PRPER.12.020132}.

Any `experiment' in a practical course can be placed on a spectrum that ranges from the very constrained `cookbook' type to entirely independent. Several classification systems exist \cite{domin1999review, bell2005simplifying, buck2008characterizing}, that share a similar structure, identifying features such whether as the outcome, method or task are given as a means to quantify the level of inquiry in an experiment \cite{buck2008characterizing}. Introducing completely independent inquiry work at the start of a degree scheme is possible but requires careful alignment of the entire programme of study. This may be challenging given constraints in courses with large theoretical components such as physics, or where  `advanced' or independent practical activities are assumed to require an high level of prior theoretical knowledge and understanding. Although students' technical expertise may be cited as a reason for `cookbook' laboratory tasks, it is unlikely that students will master the desired skills to a level where they can be used in very different or unscripted contexts without the freedom to explore the assigned task and truly experiment, and, as we demonstrate with the simple example given here, genuine inquiry can be facilitated within very simple experiments.

In this paper, we introduce the concept of `Secret Objectives' as a means of facilitating inquiry and motivating students to develop genuine inquiry attitudes in introductory physics laboratory courses. Students are explicitly challenged to find the secret objectives in each experiment. A simple damped pendulum experiment is used to provide an example of how students can be introduced to inquiry - with experiments that may be considered to lie somewhere between inquiry and discovery type \cite{domin1999review} or between guided and structured inquiry \cite{bell2005simplifying, buck2008characterizing}, but do not confine students to explore in a particular direction - within a fairly traditional teaching laboratory course. Key features of this style of adapted experiment are the simplicity of the implementation, which can often be achieved by removing parts of `cookbook' scripts that refer to the nuances or modifying the experimental design so as to not suppress imperfections. It is also possible that current experiments can be used to create secret objectives by observing behaviours that are ignored and/or compensated for by the laboratory staff. We conclude by providing some perspective connecting the secret objectives concept introduced here to problem-based learning.

\section{Secret Objectives}

In its most general form, a Secret Objective is any inconsistency between the experimental reality and the information provided to students prior to starting work on an experiment. The example experiment described here assumes that students complete set experiments with a good amount of time, and have access to written material (the `script') that includes an idealised theoretical description, basic data analysis, and an outline of the experimental procedure. The script provides enough information for students to obtain data and perform analysis, thus ensuring that all students succeed at the experiment. By advertising secret objectives, students are challenged to critically consider their data and procedure, and encouraged to look for where the theoretical description breaks down. This is a genuine inquiry activity and training in research skills that can be introduced at the very start of undergraduate studies due to its potential application to even the simplest experiments and the inbuilt requirement of experimental success.

The basic method by which secret objectives can be introduced is straightforward: instead of designing experiments so that any deviation of the data obtained from the idealised theory is so small as to be obscured by experimental uncertainties, imperfections in the experiment that would usually be removed by design are instead designed in, but the nuances not discussed in depth in the supporting material. The most simple example is to not highlight where students would get a `wrong' answer in a cookbook experiment or be walked through the explanation of the inconsistencies in a `guided inquiry' task, but to challenge students to acknowledge and explore this rather than ignoring or explaining away the deviation. This requires a certain amount of care for two reasons: the data should be fitted approximately by the idealised theory within some range, and there must be clear possible sources of discrepancy with unambiguous and testable experimental signatures. 

Not all secret objectives need to be known by the teaching team before the start of the experiment, but there should be at least one that students might be reasonably expected to find; further secret objectives are likely to be found by students investigating the one they identify first. It is important to acknowledge that secret objectives are personal and that the teaching staff only know some of them to discourage students from believing that there is a correct answer - or that there is a particular point that the demonstrator team are expecting them to find and investigate. Therefore, one important implementation point is that the course assessment should not be on experiment completion or similar, but focus on skill acquisition. For example, in the first year undergraduate laboratories at UCL where secret objectives have been developed, the assessment focuses on keeping a laboratory notebook and critical examination and exploration of the experiment.

\section{Secret objectives and a damped pendulum}

\begin{figure*}[t!]
\includegraphics[width=0.6\textwidth]{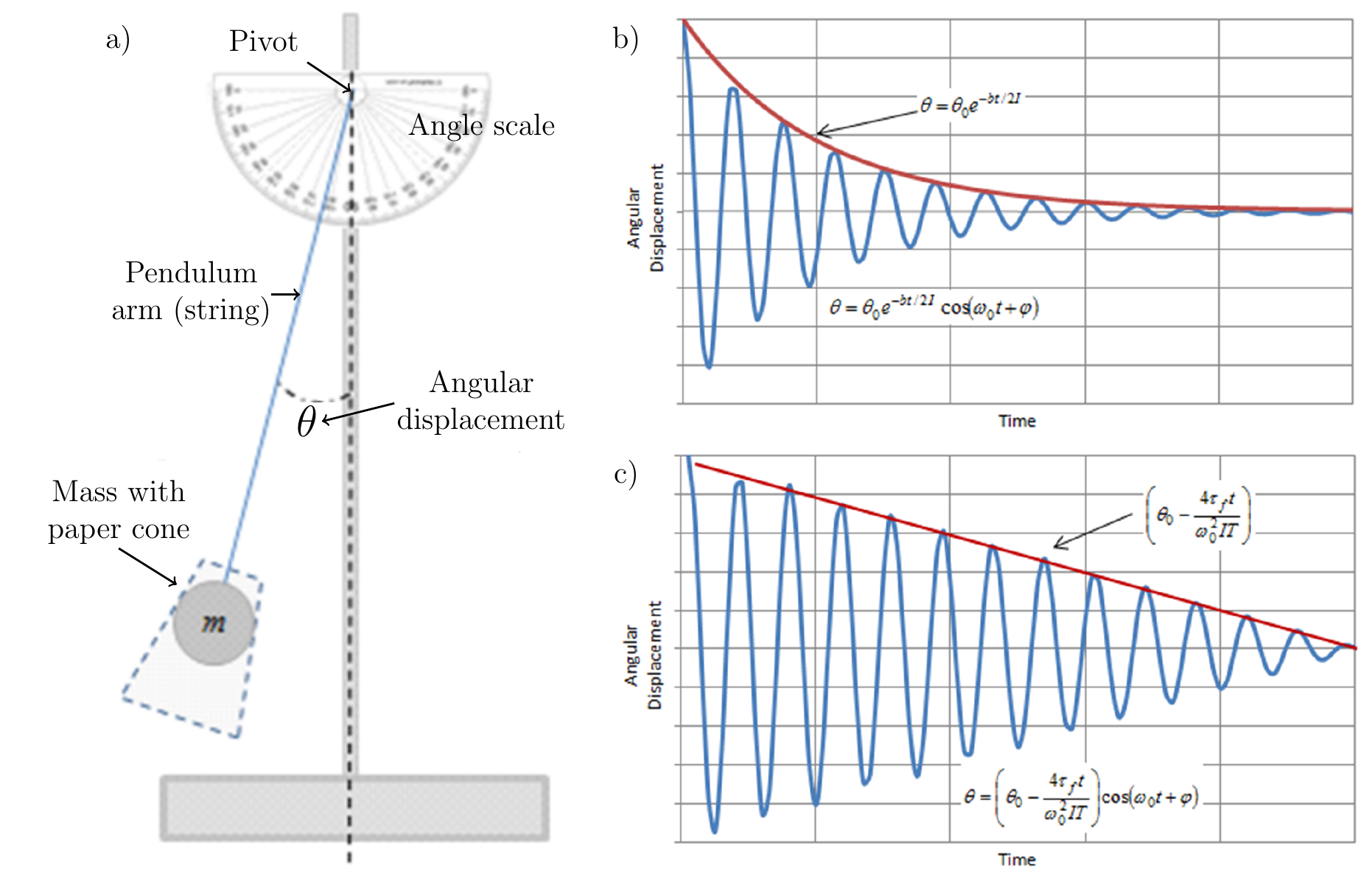}
\caption{a) Diagram a simple pendulum system with the massive bob and damping cone with angular displacement $\theta$. b) Exponential amplitude decay as results from viscous damping from the cone; c) Linear amplitude decay as consistent with dry (Coulomb) damping. \label{fig:setup}}
\end{figure*}

We use the example of a simple damped pendulum experiment (sketched in Fig. \ref{fig:setup} a) to demonstrate how secret objectives can be introduced. In the typical set up of the experiment, a bob of mass $m$ is attached to a pivot by a string is released from an angle $\theta$ away from the vertical. Attaching a paper cone to the bob introduces viscous damping, which leads to an exponential decay of the amplitude of the oscillations, as sketched in Fig. \ref{fig:setup} b \cite{BasicPhysicsTextbook-Physics}. This is the damping term most commonly described in textbooks and lecture courses. However, another type of damping, dry (Coulomb) damping, which leads to a linear decay of the amplitude of the oscillations \cite{EJP.20.85_VarDampings}, can be introduced by making the pivot imperfect. The resulting amplitude decay is sketched in Fig. \ref{fig:setup} c. 

One assumption in the description of an idealised pendulum with viscous damage is that the pivot is frictionless. The attachment of the pendulum string to the support rod is most simply achieved by threading the string through a hole in the support rod; by providing a means of securing the string, the length can be controlled easily. However, if the pendulum string is looped around the support rod and fixed in place with a knot on the underside, there is the potential for additional dry (Coulomb) damping due to friction as the string fibres rub against each other and the support rod during oscillation.

Students might be required to identify which damping mechanism is dominant when the cone is added to the pendulum bob as in Fig. \ref{fig:setup}, by comparing the data collected with and without the cone. In a demonstration type experiment, theoretical descriptions of both types of damping would be included in the script and the data obtained under different experimental conditions explained as relating to one of the two damping mechanisms with the origins given. If the script is read before or followed during the experiment is started, all students are doing is verifying a known outcome, acquiring little or no new knowledge or understanding in the process. Secret objectives can be introduced by, for example, omitting the details of the damping sources that may be present in the system. Depending on their prior theoretical knowledge, students may only be expecting viscous damping from the cone. This can mean that if Coulomb damping dominates, achieved by setting up the experiments with a messy knot at the attachment to the pivot, students will get an unexpected result which they may try to ignore (if they have a strong pre-conceived idea of what they expect to get) or they may even attempt to somehow adjust the measurement data to fit what they think should happen.

Once students have identified that the data does not agree with a model of viscous damping, student investigations could be expected to proceed in one of two obvious ways from this point, although these should not be considered restrictive, and reasonable (that is clearly defined and realisable) student investigations in any direction, even if likely to be fruitless, should be encouraged. Indeed, students should be given permission to fail, with their assessment based on what they have done rather than the completion of specific, predefined tasks that they have been told to do. One possibility is that students accept that the amplitude decay does not proceed as expected from viscous damping, perhaps removing the cone, and therefore look for an alternative theoretical description. Another possibility is that students recognise that there is friction from the attachment of the pendulum string and change the attachment accordingly, restoring the near perfect system that occurs in the description of viscous damping. In both cases, students have found a secret objective. An important task of staff supporting the experimental investigations is to allow the students to decide what the problem actually is \textit{and} what they want to do about it. Students first identify a problem with the experiment, and then inquire into its nature and the implications.

The challenge of finding the secret objective by hinting towards missing information or non-standard results is designed to make the experience of doing the experiment more enjoyable by introducing a real element of the unknown and allowing students freedom to modify the experiments accordingly. In addition, a variety of other issues or suggested avenues of investigation can be brought to students' attention, either via inclusion in the script or by questions posed directly to students, perhaps once they think they are `finished'. 

This may be the first time that students have been put into a position where they can explore and modify experiments beyond the basic set-up that they are supplied with. The missing details are important as they give students permission to deviate from the laboratory script. This fosters a spirit of investigation that should be encouraged at every opportunity, particularly with the demand from employers for analytical and problem solving skills. Further, students can also investigate the mathematical modelling of the system by producing a model which they then use to predict the performance of the pendulum under different conditions, highlighting the link between theoretical and experimental studies within physics.

Overall, any experiment designed with secret objective can be particularly useful to introduce the importance of observation during experiments and can be utilised to open a discussion about the need to understand and test the limits of a theoretical model, as well as the experimental set up. Secret objectives are also found the minute students start considering the role and validity of any assumptions they are making or using in analysing the data obtained. In pendulum experiments, this is present in the `small-angle' approximation used in deriving the equations that describe the pendulum motion \cite{BasicPhysicsTextbook-Physics}, regardless of whether or not the point has been discussed in detail in the script.

The range of aspects that students can explore means that they can discuss the experiment and their own investigations on a larger scale, perhaps formed of several experimental teams of 2-4 students. This helps to show students the strengths and weaknesses of theoretical descriptions as well as fostering a culture of sharing ideas as well as providing experience in communicating scientific investigations.

By being open about the presence of inconsistencies, students are prompted to be more critical about the data they obtain and its interpretation. They should see these deliberate omissions as challenges that develop their investigative skills as well as promoting ownership of the experiments via personalised modifications. An observed limitation on the success of such experiments is that assessment must be aligned to recognise and reward the inquiry behaviours rather than emphasising task completion or record keeping \cite{EJP.40.015702}. The damped pendulum example discussed here is a particularly simple example that requires little, if any, specialised equipment not already available in even high-school science departments. Other topics for which an inexpensive secret objectives experiment could be easily devised include investigations into the conservation of momentum, friction and air resistance in projectile motion, and resonance in operational amplifiers to suggest just a few. In addition, experiments that are difficult to set up in a way that the problems with them are removed, and thus traditionally require detailed discussions of the imperfections, naturally contain secret objectives. Recognising and understanding the imperfections becomes, with the advertisement of the existence of secret objectives, a part of the laboratory work without modifying the experiments.

\section{Conclusions and perspective}

We have introduced the concept of `secret objectives' advertising the presence of discrepancies between predicted and realised experimental data as a means of challenging and motivating students to engage in genuine inquiry behaviours in a laboratory course. Secret objectives allow the structure of common teaching experiments to be maintained, even to the outline of the method and the approximate agreement between experimental data and idealised theory, which both guarantees students' success and simplifies implementation. The crucial factor in successful implementation is time, with several laboratory sessions devoted to each experiment being ideal. The emphasis of the laboratory work is thus moved onto creating habits of critical analysis of the data and development of understanding rather than on the completion of set laboratory tasks.

Outside of the unique environment of teaching laboratories, problem-based learning (PBL) is a popular teaching and learning activity \cite{TeachHE.5.345, PBLFoundations}. In most PBL scenarios, the problem is usually stated and the students tasked to come up with an appropriate solution or diagnosis \cite{PBLFoundations}. PBL is not common in physics, although it can lead to improved conceptual understanding \cite{JSciEdTech.19.266, IntEdJ.6.430}. In particular, there are known difficulties in the use of PBL in basic science and engineering courses \cite{TeachHE.5.345} and its application to theoretical work with a pre-defined outcome may not be optimal \cite{PBLFoundations}.

In pedagogical terms, the secret objectives concept bridges the boundary between inquiry style laboratories and PBL: when a secret objective - i.e.: `problem' with the experiment - is found, the students have created the problem that might be considered the starting point for a more traditional PBL activity. Thus, the secret objectives concept, as well as being useful on its own, can also be used to quite naturally align laboratory courses where students may struggle with a lack of structure to an ethos that emphasises PBL, or even to effectively facilitate PBL in a consistent manner without disturbing more traditional lecture courses. Importantly, secret objectives explicitly do not define a correct solution from the outset, and are therefore free of an instructor assumption that can occur in more strictly defined PBL scenarios \cite{PBLFoundations}.

\bibliographystyle{unsrt}
\bibliography{teachingbibliography_v2,Textbooks}

\end{document}